\documentclass[aps,pra,nofootinbib,twocolumn]{revtex4-2}

\usepackage{amsmath,amssymb,amsthm,mathtools,bm,physics}
\usepackage{graphicx}
\usepackage[colorlinks=true,citecolor=blue,linkcolor=blue,urlcolor=blue]{hyperref}
\usepackage{orcidlink}
\usepackage[normalem]{ulem}
\usepackage{comment}

\newcommand{\lM}{\lambda_{\rm max}}
\newcommand{\lm}{\lambda_{\rm min}}
\newcommand{\lSm}{\lambda_{\rm Smin}}
\newcommand{\I}{\mathop{\mathbb{I}}\nolimits}
\newcommand{\Exp}[1]{\langle #1 \rangle}
\newcommand{\zExp}[1]{\hat{#1}}

\theoremstyle{plain}
\newtheorem{theorem}{Theorem}

\newtheorem{proposition}{Proposition}
\newtheorem{corollary}{Corollary}
\theoremstyle{definition}

\theoremstyle{remark}
\newtheorem{remark}{Remark}

\newcommand{\gk}[1]{{\textcolor{black}{#1}}}

\newenvironment{gkblock}
{\color{black}}
{}

\begin{document}

\title{Tight Generalization of Robertson-Type Uncertainty Relations}

\author{Gen Kimura\,\orcidlink{0000-0003-4288-2024}}
\affiliation{
Graduate School of Information Sciences, Tohoku University, Sendai 980-8579, Japan
}
\email{gen.kimura.quant@gmail.com}
\author{Aina Mayumi\,\orcidlink{0009-0007-0956-1620}}%
\author{Haruki Yamashita\,\orcidlink{0009-0002-2439-0317}}%
\email{haruki.yamashita1367@gmail.com}
\affiliation{Graduate School of Engineering and Science, Shibaura Institute of Technology, 307~Fukasaku, Minuma-ku, Saitama 337-0003, Japan}

\date{\today}

\begin{abstract}
We establish the tightest possible Robertson-type preparation uncertainty relation, which explicitly depends on the eigenvalues of the quantum state. The conventional constant \( \tfrac{1}{4} \) is replaced by the state-dependent coefficient
\[
\frac{(\lambda_{\max} + \lambda_{\min})^2}{4(\lambda_{\max} - \lambda_{\min})^2},
\]
where \( \lambda_{\max} \) and \( \lambda_{\min} \) denote the largest and smallest eigenvalues of the density operator \( \rho \), respectively. 
We prove that this coefficient is optimal among all Robertson-type
generalizations and does not admit further improvement.
\gk{We also obtain a complete characterization of all nontrivial equality cases.} 
Our relation becomes more pronounced as the quantum state becomes more mixed, capturing a trade-off in quantum uncertainty that the conventional Robertson's relation fails to detect. In addition, our result also provides a strict generalization of the Schr\"odinger's uncertainty relation, showing that the uncertainty trade-off is governed by the sum of the covariance term and a state-dependent improvement over the Robertson bound.
As applications, we also refine error-disturbance trade-offs by incorporating spectral information of both the system and the measuring apparatus, thereby generalizing the Arthurs-Goodman and Ozawa inequalities.
\end{abstract}


\maketitle

\section{Introduction}
The canonical uncertainty principle reveals that two non‑commuting observables cannot simultaneously assume arbitrarily sharp values \cite{Heisenberg1927,Peres1995,Wehner_2010,HilgevoordUffink2024}. 
Building on Kennard's formulation of uncertainty in terms of standard deviations for position and momentum observables~\cite{Kennard1927}, Robertson's uncertainty relation~\cite{Robertson1929} established a general trade-off relation between arbitrary observables, thereby revealing an intrinsic quantum uncertainty:
\begin{align}\label{eq:Rob}
 V_{\rho}(A)V_{\rho}(B) &\ge \frac14 \Bigl|\Exp{[A,B]}_{\rho}\Bigr|^{2}, 
\end{align}
where $A$, $B$ are observables represented by Hermitian operators, $V_\rho(X) := \Tr[(X-\langle X \rangle_\rho)^2\rho]$ and $\langle X \rangle_\rho := \Tr [X\rho]$ denote the variance of and the expectation value of $X$ in the quantum state $\rho$, and $[A,B]:= AB-BA$ denotes the commutator. 
The relation was soon refined by Schr\"odinger~\cite{Schrodinger1930}, who added to the Robertson bound a quantum covariance term
\(
\mathrm{Cov}_\rho(A, B) \coloneqq \frac{1}{2} \langle \{ A, B \} \rangle_\rho - \langle A \rangle_\rho \langle B \rangle_\rho,
\)
with the anticommutator \( \{A, B\} \coloneqq AB + BA \):
\begin{equation}\label{eq:SUR}
V_\rho(A)\, V_\rho(B) \ge \frac{1}{4} \Bigl|\Exp{[A,B]}_{\rho}\Bigr|^{2} + \mathrm{Cov}_\rho(A, B)^2.
\end{equation}
Following the pioneering work of Robertson and its refinement by Schr\"odinger, many developments have been made on preparation-type uncertainty relations.
Maccone and Pati \cite{Maccone} proposed an improvement to Robertson's relation using the sum of variances \cite{Yichen}, which has led to further developments \cite{Xiao:17, SONG20162925, Fan2020}.
On the other hand, Luo \cite{Luo2} and Park \cite{Park2005} independently proposed improvements for mixed states using the Wigner–Yanase skew information \cite{wigner1963}, and several related studies have also been conducted \cite{LZ2004, Yanagi_2005, LZ2005, Kosaki_2005, Li_2009, FURUICHI2009179, Yanagi_2010, Chen_2016}. 
Another refinement of Robertson's relation~\eqref{eq:Rob} for mixed state was also proposed by Nagaoka and Hayashi ~\cite{Hayashi2006QI,Nagaoka2005CRbound}, while several studies have characterized  uncertainty regions \cite{Li2015, Busch2019, Zhang2021, math4010008}.
Besides variance, formulations using other measures—such as entropy \cite{Deutsch, Maassen, Kraus_1987, Berta_2010, RevModPhys.89.015002, HallRenyi}, maximum probabilities \cite{Landau1961, MiyaderaImai}, Fisher information \cite{FSG2015, CG2022, TGFF2022}, and quantum coherence \cite{SPB2016, Liu2016, Plenio2, PhysRevA.96.032313, Rastegin2017, Luo_2019}—have also been developed.

In a recent work~\cite{MKHC2},  we considered the following type of uncertainty relation:
\begin{equation}\label{I}
V_\rho(A)\, V_\rho(B) \ge c(\rho) 
\gk{\Exp{\bigl|[A,B]\bigr|^2}_{\rho}},
\end{equation}
where \( |X| := \sqrt{X^\dagger X} \) is the matrix absolute value, and \( c(\rho) \) is a positive constant that may depend on the state \( \rho \). 
We showed that
\(
c(\rho) = \lm/(2\lM^2)
\)
yields a valid general uncertainty relation, where \( \lm \) and \( \lM \) denote the smallest and largest eigenvalues of \( \rho \), respectively, while the optimal bound was conjectured to be
\begin{equation}\label{copt}
c_{\rm opt}(\rho) \gk{:=\inf_{\substack{
A=A^\dagger,\;B=B^\dagger\\
\Exp{\bigl|[A,B]\bigr|^2}_{\rho}\neq 0
}}
\frac{V_\rho(A)V_\rho(B)}
{\Exp{\bigl|[A,B]\bigr|^2}_{\rho}}}= \frac{\lm \lSm}{\lm + \lSm},
\end{equation}
where \( \lSm \) is the second-smallest eigenvalue of \( \rho \). 
Recently, in Ref.~\cite{KMOLC}, we not only provided a rigorous proof of this result, but also further strengthened the Schr\"odinger uncertainty relation by incorporating the additional bound~\eqref{copt} into the original expression~\eqref{eq:SUR}:
\begin{align}\label{eq:UR_2}
V_\rho(A)\, V_\rho(B) \ge\ \frac{1}{4} \left| \Exp{[A, B]}_\rho \right|^2 + \mathrm{Cov}_\rho(A, B)^2 \nonumber \\
\quad + \frac{\lm \lSm}{\lm + \lSm} \gk{\Exp{\bigl|[A,B]\bigr|^2}_{\rho}}.
\end{align}
The last term reveals a previously overlooked quantum contribution to the uncertainty trade-off, whose significance increases as the quantum state becomes more mixed. 

Thus, the uncertainty relation of the form~\eqref{I}, based on the commutator norm dependent on \(\rho\), has been established in its optimal form. 
It is, however, still of interest to pursue a direct refinement of the Robertson-type relation: Namely, we consider the problem of determining the largest possible positive coefficient \( c'(\rho) \) such that the following uncertainty relation holds:
\begin{equation}\label{II}
V_\rho(A) V_\rho(B) \geq c'(\rho) \Bigl|\Exp{[A,B]}_{\rho}\Bigr|^{2}.
\end{equation}
Note that the case \( c'(\rho) = 1/4 \) corresponds to the original Robertson's relation \eqref{eq:Rob}. 
Somewhat surprisingly, however, this bound is not yet optimal and can be further improved.  
We show that the optimal bound is determined by the minimum and maximum eigenvalues of the density operator, and is given by
\begin{equation}\label{URIIc}
c'_{\rm opt}(\rho) \gk{:=\inf_{\substack{
A=A^\dagger,\;B=B^\dagger\\
\bigl|\Exp{[A,B]}_{\rho}\bigr|^2\neq 0
}}
\frac{V_\rho(A)V_\rho(B)}
{\bigl|\Exp{[A,B]}_{\rho}\bigr|^2}} = \frac{(\lM + \lm)^2}{4(\lM - \lm)^2},
\end{equation}
leading to a direct generalization of Robertson's
uncertainty relation:
\begin{equation}\label{URII}
V_\rho(A)\, V_\rho(B) \geq \frac{(\lM + \lm)^2}{4(\lM - \lm)^2} \Bigl|\Exp{[A,B]}_{\rho}\Bigr|^{2}. 
\end{equation}
This relation yields a strictly sharper bound than the Robertson bound \eqref{eq:Rob} for all faithful states, i.e., when \( \lambda_{\min} > 0 \). (\gk{The exceptional case $\lambda_{\max}=\lambda_{\min}$, corresponding to the maximally mixed state, requires separate treatment and is discussed in detail in Sec.~\ref{Sec:2}}.)
Interestingly, while the smallest and the second smallest eigenvalues of $\rho$ determine the optimal bound \eqref{copt} in the case of the commutator norm, the optimal bound of the Robertson type is determined by the smallest and the largest eigenvalues of $\rho$.
Moreover, using a similar technique introduced in Ref.~\cite{Wu2010}, uncertainty relation \eqref{URII} can be further generalized to
\begin{equation}\label{URII-2}
V_{\rho}(A) V_{\rho}(B) \geq \frac{(\lM + \lm)^2}{4(\lM - \lm)^2} \left|\langle [A, B] \rangle_\rho \right|^2  + {\rm Cov}_\rho(A, B)^2,
\end{equation}
which provides another refinement of the Schr\"odinger's uncertainty relation.

While preparation-type uncertainty relations have been the focus so far, measurement-type counterparts also play a central role in the understanding of quantum measurement, as they explicitly characterize the trade-off between measurement precision and the disturbance it inevitably induces \cite{ArthursKelly1965,ArthursGoodman1988,Ozawa2003,Ozawa2004,Werner2004,Miyadera2008,BuschLahtiWerner2013Proof,BuschLahtiWerner2014MeasurementUR,Buscemi2014InfoNoiseDist,Ozawa2014MixedStates}.
As an application of our uncertainty relation, we also obtain generalizations of measurement-type uncertainty relations, including the Arthurs–Goodman inequality and Ozawa's inequality \cite{ArthursGoodman1988,Ozawa2003,Ozawa2004}.

The paper is organized as follows.
In Sec.~\ref{Sec:2}, we discuss a Robertson-type uncertainty relation and derive its tight generalization.  
In Sec.~\ref{Sec:3}, we use this result to obtain a tight generalization of Schr\"odinger's relation.
In Sec.~\ref{Sec:equality}, we characterize the equality conditions for both optimized relations and compare their tightness with the conventional Robertson and Schr\"odinger bounds, including a geometric interpretation for qubits.
In Sec.~\ref{Sec:4}, we compare our generalized bounds with the original Robertson and Schr\"odinger relations in the qubit case by averaging the bounds over all pairs of observables.
In Sec.~\ref{Sec:5}, we apply our generalized uncertainty relation to measurement-type uncertainty relations and obtain strengthened versions of the Arthurs–Goodman and Ozawa inequalities.
Finally, Sec.~\ref{Sec:6} summarizes our results.

\section{Generalization of Robertson's relation}\label{Sec:2}

In this section, we show that the optimal bound \( c'(\rho) \) for uncertainty relations of the form~\eqref{II} is given by~\eqref{URIIc}. This is stated in the following theorem:
\begin{theorem}\label{thm1}
Let \( A \) and \( B \) be arbitrary quantum observables, and let \( \rho \) be a quantum state that is not maximally mixed. Then the following uncertainty relation holds:
\begin{equation}\label{main:UR}
V_\rho(A)\, V_\rho(B) \ge 
    \frac{(\lM + \lm)^2}{4(\lM - \lm)^2} \left| \Exp{[A,B]}_{\rho} \right|^{2},
\end{equation}
where \( \lm \) and \( \lM \) denote the smallest and largest eigenvalues of \( \rho \), respectively. 
\gk{Moreover, the coefficient on the right-hand side of
Eq.~\eqref{main:UR} is optimal. 
That is, for each fixed $\rho$, it cannot be replaced
by any larger constant while preserving the inequality for all
observables $A$ and $B$.}
\end{theorem}
{\it Proof.} Note that for non-faithful states with \( \lambda_{\min} = 0 \), the bound~\eqref{URIIc} reduces to \( 1/4 \), thus recovering Robertson’s relation. This observation also applies to infinite-dimensional systems, where \( \lambda_{\min} \) asymptotically approaches zero. Therefore, it suffices to prove~\eqref{main:UR} for finite-dimensional systems and faithful states with \( \lambda_{\min} > 0 \). Let \( d \) be the dimension of the Hilbert space; then, with an appropriate choice of basis, \( A \) and \( B \) can be represented by \( d \times d \) Hermitian matrices. In what follows, we work in the basis \( \{ \ket{i} \}_{i=1}^d \) that diagonalizes \( \rho \), and denote the matrix elements of \( A \) and \( B \) by \( a_{ij} := \braket{i}{A|j} \) and \( b_{ij} := \braket{i}{B | j} \), respectively, for \( i,j = 1, \ldots, d \). The eigenvalues of \( \rho \) are denoted by \( \lambda_i \ (i = 1, \ldots, d) \), and are arranged in ascending order:
\[
0 < \lm := \lambda_1 \le \lambda_2 \le \cdots \le \lambda_d =: \lM.
\]
Note that \( \lambda_d > \lambda_1 \) since \( \rho \) is not the maximally mixed state.

Introducing the centered observable
\begin{equation}\label{shift}
\zExp{X} := X - \langle X \rangle_\rho \I,
\end{equation}
the variance of \( X \) can be written as
\[
V_\rho(X) = \| \zExp{X} \|^2_\rho,
\]
where $\|X\|_\rho:= \sqrt{\Tr X^\dagger X \rho}$ denotes the \(\rho\)-weighted Frobenius norm.
Since \( [\zExp{A}, \zExp{B}] = [A, B] \), the inequality~\eqref{main:UR} can be reformulated as
\begin{equation}\label{URnorm}
    \| \zExp{A} \|^2_\rho\, \| \zExp{B} \|^2_\rho \geq \frac{(\lambda_d + \lambda_1)^2}{4(\lambda_d - \lambda_1)^2} \left| \Exp{[\zExp{A}, \zExp{B}]}_{\rho} \right|^2.
\end{equation}
In the following, we prove that for arbitrary Hermitian matrices $A,B$,
\begin{equation}\label{ineq:rhonorm}
     \| A \|^2_\rho\, \| B \|^2_\rho 
     \geq \frac{(\lambda_d + \lambda_1)^2}{4(\lambda_d - \lambda_1)^2}
     \left| \Exp{[A, B]}_{\rho} \right|^2,
\end{equation}
so that the relation \eqref{URnorm} follows by replacing \(A\) with \(\hat A\) and \(B\) with \(\hat B\). 

To show the inequality \eqref{ineq:rhonorm}, we decompose the matrices \( A \) and \( B \) into their diagonal and off-diagonal components (with respect to the basis of $\rho$) as
\begin{equation}
A = A_d + A_n, \quad B = B_d + B_n,
\end{equation}
where \( A_d \) and \( B_d \) are diagonal matrices, and \( A_n \), \( B_n \) contain the off-diagonal parts. Note that \( \Tr([A, B]\rho) = \Tr([A_n, B_n]\rho) \). It is also straightforward to verify that the diagonal and off-diagonal components are orthogonal with respect to the inner product
\begin{equation}\label{rhoHS}
\langle X, Y \rangle_\rho := \Tr(X^\dagger Y \rho),
\end{equation}
defined by a faithful state \( \rho \). Therefore, we have
\begin{equation}\label{eq:diagonal-reduction}
\begin{aligned}
\|A\|^2_\rho
&=
\|A_d\|^2_\rho+\|A_n\|^2_\rho
\ge
\|A_n\|^2_\rho,\\
\|B\|^2_\rho
&=
\|B_d\|^2_\rho+\|B_n\|^2_\rho
\ge
\|B_n\|^2_\rho.
\end{aligned}
\end{equation}
Hence, to prove~\eqref{main:UR}, it suffices to show \gk{\eqref{ineq:rhonorm}} 
for Hermitian matrices with zero diagonal entries: $a_{ii} = b_{ii} = 0 \ (i = 1,\ldots, d)$. 

A direct computation then yields
\begin{equation}\label{eq:ABrnorm}
\| A \|_{\rho}^2 = \sum_{i<j} (\lambda_i + \lambda_j) |a_{ij}|^2,\  \| B \|_{\rho}^2 = \sum_{i<j} (\lambda_i + \lambda_j) |b_{ij}|^2,
\end{equation}
and 
\begin{align}\label{eq:commutator-elements}
\Bigl| \Exp{[A,B]}_{\rho}\Bigr|^2
&=
\Bigl| \sum_{i<j} (\lambda_j-\lambda_i)\Bigl(a_{ji}{b_{ij}}-a_{ij}{b_{ji}}\Bigr)\Bigr|^2 \nonumber \\
&=
4\Bigl| \sum_{i<j} (\lambda_j-\lambda_i)
\Im \Bigl(a_{ij}\overline{\gk{b_{ij}}}\Bigr)\Bigr|^2.
\end{align}
From the last expression, it follows 
\begin{equation}\label{eq:wq4}
\frac{1}{4(\lambda_d-\lambda_1)^2}\Bigl|\langle[A,B]\rangle_\rho\Bigr|^2
\le \Biggl|\sum_{i<j} \frac{\lambda_j-\lambda_i}{\lambda_d-\lambda_1}a_{ij}\overline{b_{ij}}\Biggr|^2. 
\end{equation}
By the Schwarz inequality, this is further bounded by 
\begin{equation}\label{eq:wq1}
\Biggl(\sum_{i<j}\frac{\lambda_j-\lambda_i}{\lambda_d-\lambda_1}|a_{ij}|^2\Biggr)
\Biggl(\sum_{i<j}\frac{\lambda_j-\lambda_i}{\lambda_d-\lambda_1}|b_{ij}|^2\Biggr). 
\end{equation}
Now, it is straightforward to show that\footnote{
The inequality~\eqref{eq:wq3} is equivalent to
\begin{equation*}
(\lambda_j - \lambda_i)(\lambda_1 + \lambda_d) \le (\lambda_i + \lambda_j)(\lambda_d - \lambda_1),
\end{equation*}
which, upon expansion and simplification, reduces to
\begin{equation*}
\lambda_i \lambda_d \ge \lambda_j \lambda_1.
\end{equation*}
This holds trivially since \( \lambda_d \ge \lambda_j \) and \( \lambda_i \ge \lambda_1 \), due to the ordering of the eigenvalues. }, for each pair \( i < j \),
\begin{equation}\label{eq:wq3}
\frac{\lambda_j - \lambda_i}{\lambda_d - \lambda_1} \le \frac{\lambda_i + \lambda_j}{\lambda_1 + \lambda_d}.
\end{equation}
Applying this inequality to Eq.~\eqref{eq:wq1} and using Eq.~\eqref{eq:ABrnorm}, we obtain Eq.~\eqref{ineq:rhonorm}, thereby proving the uncertainty relation~\eqref{main:UR}.

To demonstrate the tightness of the inequality, it suffices to show that for any density matrix~\(\rho\) satisfying $\lambda_{\max}>\lambda_{\min}$, there exist observables $A$ and $B$ for which equality holds with both sides strictly positive.
Indeed, if one were to increase $c'(\rho)$ in \eqref{II} further, the inequality would no longer hold for these observables.  
One such pair is given below, whose supports lie in the eigenspaces of the density operator \(\rho\) corresponding to \(\lm\) and \(\lM\)\footnote{Note that the example \eqref{eq:ex} consists of observables with vanishing expectation values, so that \(\hat{A}=A,\hat{B}=B\). 
Therefore, the equality in the norm inequality \eqref{ineq:rhonorm} (for arbitrary Hermitian operators \(A\) and \(B\)) directly corresponds to equality in the uncertainty relation \eqref{main:UR}.}:
\begin{align}\label{eq:ex}
A&=\ketbra{d}{1}+\ketbra{1}{d},
\quad \nonumber \\ 
B&=i\bigl(\ketbra{d}{1}-\ketbra{1}{d}\bigr).
\end{align}
This completes the proof. 
\qed

In Sec.~\ref{Sec:equality}, Proposition~\ref{prop:equality-gR} further provides a complete characterization of all observables that attain equality in the generalized Robertson relation.

As already mentioned in the introduction, one cannot use \eqref{main:UR} for the maximally mixed state \( \rho_{\max} = \I/d \) for a \( d \)-level system, since the eigenvalues satisfy \( \lM = \lm = 1/d \), and hence the bound in \eqref{main:UR} formally diverges. 
To see whether we have a non-trivial bound for the maximally mixed state, consider the sequence of faithful states $\rho_n \ (n = d+1,d+2,\ldots)$ with eigenvalues 
\begin{equation}
\lambda_1 = \frac{1}{d} - \frac{1}{n}, \lambda_i = \frac{1}{d} + \frac{1}{n} \cdot \frac{1}{d-1}\quad(i \geq 2), 
\end{equation}
with the common eigenvectors $\{\ket{i}\}_{i=1}^d$. 
Clearly, $\rho_{\max} = \lim_{n \to \infty} \rho_n$ for any basis $\{\ket{i}\}_{i=1}^d$. 

By direct computation, one can estimate the right-hand side of~\eqref{main:UR} as
\[
\frac{\left( \frac{2}{d} + \frac{2-d}{n(d-1)}\right)^2}{4} \cdot
\left| \sum_{j = 2}^{d} \left( a_{1j} \overline{b}_{1j} - \overline{a}_{1j} b_{1j} \right) \right|^2.
\]
Taking the limit \( n \to \infty \), this expression converges to
\[
\frac{1}{d^2} \left| \sum_{j = 2}^{d} \left( a_{1j} \overline{b}_{1j} - \overline{a}_{1j} b_{1j} \right) \right|^2,
\]
which can be rewritten as
\begin{align*}
&\frac{1}{d^2} \left| \sum_{j = 1}^{d} \left( a_{1j} \overline{b}_{1j} - \overline{a}_{1j} b_{1j} \right) \right|^2 \\
&= \frac{1}{d^2} \left| \sum_{j=1}^{d} \left( \langle 1 | A | j \rangle \langle j | B | 1 \rangle - \langle 1 | B | j \rangle \langle j | A | 1 \rangle \right) \right|^2\nonumber \\
&= \frac{1}{d^2} \Bigl| \langle 1 | [A, B] | 1 \rangle \Bigr|^2
\end{align*}
where in the final equality we used the completeness relation \( \I = \sum_j \ketbra{j}{j} \).

Since the choice of basis for diagonalizing the density operator \( \rho \) was arbitrary, the vector \( \ket{1} \) can be replaced by any unit vector $\ket{\psi}$. Therefore, it follows from \eqref{main:UR} that 
\begin{equation}
V_{\rho_{\max}}(A)\, V_{\rho_{\max}}(B) \geq \frac{1}{d^2} \Bigl|\braket{\psi}{[A,B]\psi}\Bigr|^2.     
\end{equation}

It is well known (see, e.g., Ref.~\cite{BhatiaMatrixAnalysis}) that, for any anti-Hermitian operator, the operator norm \( \| X \|_{\rm{op}} := \max_{\ket{\psi} \neq 0} \|X \psi\|/\|\psi\|\) satisfies
\[
\|X\|_{\mathrm{op}} = \max_{\ket{\psi}} \left| \braket{\psi}{X | \psi} \right|. 
\]
Therefore, in the case of the maximally mixed state, we obtain 
\begin{equation}\label{URMM}
V_{\rho_{\max}}(A) V_{\rho_{\max}}(B) \geq \frac{1}{d^2} \| [A,B] \|_{\rm{op}}^2. 
\end{equation}
Note that, whereas Robertson's relation becomes trivial in the case of the maximally mixed state, the bound \eqref{URMM} remains nontrivial for any pair of non-commuting observables.

\section{Generalization of Schr\"odinger's relation}\label{Sec:3} 
Interestingly, our relation~\eqref{main:UR} not only generalizes Robertson's relation, but also yields a generalization of Schr\"odinger's relation:
\begin{theorem}Let \( A \) and \( B \) be arbitrary quantum observables, and let \( \rho \) be a quantum state that is not maximally mixed. Then the following uncertainty relation holds:
\begin{equation}\label{main:UR2}
V_\rho(A) \, V_\rho(B) 
\ge \frac{(\lM + \lm)^2}{4(\lM - \lm)^2} 
    \left| \langle [A, B] \rangle_\rho \right|^2  + {\rm Cov}_\rho(A, B)^2. 
\end{equation}
\gk{Moreover, the coefficient of the first term on the right-hand side is optimal in the same sense as in Theorem~\ref{thm1}.}
\end{theorem}
The following proof is inspired by the technique used in Ref.~\cite{Wu2010}, which is generalized here in a state-dependent manner.

{\it Proof}. Let \( \alpha, \beta \neq 0 \in \mathbb{R} \), and define Hermitian operators 
\begin{equation}\label{PQ}
P := \alpha A - \beta B,\ Q := \alpha A + \beta B.
\end{equation}
Using the identity
\[
[P, Q] = 2 \alpha \beta [A, B], 
\]
we have 
\[
|\langle [A, B] \rangle_\rho|^2 = \frac{1}{4 \alpha^2 \beta^2} |\langle [P, Q] \rangle_\rho|^2. 
\]
Applying inequality~\eqref{ineq:rhonorm} to the operators \( P \) and \( Q \), we obtain
\begin{equation}\label{URnormPQ}
\left| \langle [A, B] \rangle_\rho \right|^2 
\leq \frac{(\lambda_{\max} - \lambda_{\min})^2}{\alpha^2 \beta^2 (\lambda_{\max} + \lambda_{\min})^2} \, 
\| P \|^2_\rho \, \| Q \|^2_\rho.
\end{equation}
Substituting the definitions of \( P = \alpha A - \beta B \) and \( Q = \alpha A + \beta B \), and expanding the norms, we find
\[
\| P \|^2_\rho \, \| Q \|^2_\rho 
= \left[ \left( \alpha^2 \| A \|_\rho^2 + \beta^2 \| B \|_\rho^2 \right)^2 
- \left( 2 \alpha \beta \operatorname{Re} \Tr(A B \rho) \right)^2 \right].
\]

Now, choosing \( \alpha \) and \( \beta \) such that \( \alpha^2 \| A \|_\rho^2 = \beta^2 \| B \|_\rho^2 \), i.e., the equality condition for the arithmetic--geometric mean inequality,
\(
\left( \alpha^2 \| A \|_\rho^2 + \beta^2 \| B \|_\rho^2 \right)^2 
\ge 4 \alpha^2 \beta^2 \| A \|_\rho^2 \| B \|_\rho^2,
\)
we obtain
\[
\| P \|^2_\rho \, \| Q \|^2_\rho 
= 4 \alpha^2 \beta^2 \left( \| A \|_\rho^2 \| B \|_\rho^2 - \left( \operatorname{Re} \Tr(A B \rho) \right)^2 \right).
\]
Using this in~\eqref{URnormPQ}, and the identity \( \operatorname{Re} \Tr(A B \rho) = \frac{\langle \{ A, B \} \rangle_\rho}{2} \), we conclude that
\begin{align}\label{URnormPQ2}
\left| \langle [A, B] \rangle_\rho \right|^2 
&\leq \frac{4 (\lambda_{\max} - \lambda_{\min})^2}{(\lambda_{\max} + \lambda_{\min})^2} \nonumber \\
&\quad \times \left[ \| A \|_\rho^2 \| B \|_\rho^2 
- \left( \frac{\langle \{ A, B \} \rangle_\rho}{2} \right)^2 \right].
\end{align}
Finally, by applying the centered observables \( \hat{A} = A - \langle A \rangle_\rho \I \) and \( \hat{B} = B - \langle B \rangle_\rho \I \) to~\eqref{URnormPQ2}, and noting again that
\(
V_\rho(A) = \| \hat{A} \|^2_\rho, \quad V_\rho(B) = \| \hat{B} \|^2_\rho, \quad [\hat{A},\hat{B}]=[A,B],
\)
as well as
\[
\mathrm{Cov}_\rho(A, B) = \frac{\langle \{ \hat{A}, \hat{B} \} \rangle_\rho}{2},
\]
we obtain the generalized Schr\"odinger-type uncertainty relation~\eqref{main:UR2}. 

\gk{Finally, the tightness of relation~\eqref{main:UR2} is established by
the same pair of observables $A$ and $B$ defined in
Eq.~\eqref{eq:ex}. Indeed, both observables have vanishing
expectation values and anticommute, i.e., $\{A,B\}=0$, and hence
${\rm Cov}_\rho(A,B)=0$. Consequently, since this pair saturates
Eq.~\eqref{main:UR}, it also saturates Eq.~\eqref{main:UR2}.}
\qed 

\gk{The complete equality conditions (Proposition~\ref{prop:equality-gS}), together with their qubit interpretation, are discussed in the next section.}

\begin{gkblock}
\section{Equality conditions and the meaning of tightness}\label{Sec:equality}

Equality conditions in uncertainty relations are often discussed in connection with minimum-uncertainty states, especially for the canonical position--momentum pair and related coherent or squeezed states~\cite{Trifonov2001,Englert2024}.
For the Kennard relation, the lower bound is state independent, so saturation naturally singles out states that attain the smallest possible uncertainty product for the fixed pair of observables.

In finite dimensions, however, the situation is different.
For fixed observables $A$ and $B$, minimizing $V_\rho(A)V_\rho(B)$ over states is generally trivial, since an eigenstate of either observable makes one of the variances vanish.
Thus, in the present finite-dimensional setting, the physically relevant role of an equality condition is not to identify a minimum-uncertainty state, but rather to determine, for a prescribed state $\rho$, which pairs of observables make a given uncertainty relation tight.
This provides a natural way to compare the conventional Robertson and Schr\"odinger relations with their state-dependent optimal refinements.

Our strict state-dependent improvements~\eqref{main:UR} and~\eqref{main:UR2} immediately imply that any pair of observables saturating the conventional Robertson or Schr\"odinger relation must satisfy $\Exp{[A,B]}_\rho=0$, since otherwise the strictly larger coefficient \(c'(\rho)>1/4\) would contradict saturation of the conventional relation. 
Moreover, for faithful states, Gudder's characterization \cite{Gudder2023} of the equality condition for the Schr\"odinger relation further implies that the centered observables are real-linearly dependent, and hence $[A,B]=0$. Indeed, with
\(
\hat A=A-\Exp{A}_\rho\I
\)
and
\(
\hat B=B-\Exp{B}_\rho\I
\),
one has
\begin{equation}
\langle \hat A,\hat B\rangle_\rho
=
{\rm Cov}_\rho(A,B)
+
\frac{1}{2}\Exp{[A,B]}_\rho .
\end{equation}
Hence, the conventional Schr\"odinger relation~\eqref{eq:SUR} is precisely the Cauchy--Schwarz inequality for $\hat A$ and $\hat B$ with respect to the $\rho$-weighted inner product.
For a faithful state, \eqref{rhoHS} defines a genuine inner product, and equality in the Cauchy--Schwarz inequality holds if and only if the two centered observables are linearly dependent.
If both variances are nonzero, Hermiticity further implies that the proportionality constant is real, so that
\begin{equation}\label{eq:standard-SR-equality}
\hat B=t\hat A,
\quad
t \neq 0 \in \mathbb{R}
\end{equation}
Consequently, $[A,B]=0$. 
Thus, for a faithful state, the conventional Schr\"odinger relation cannot be saturated by a genuinely noncommuting pair. 

The same conclusion holds a fortiori for the conventional Robertson relation.
Indeed, equality in Robertson's relation with a nonzero commutator expectation would require the covariance term in the Schr\"odinger relation to vanish and would therefore force equality in the Schr\"odinger relation as well, which is impossible for a noncommuting pair by Eq.~\eqref{eq:standard-SR-equality}.
Hence, for faithful mixed states, both conventional relations possess an unavoidable gap whenever the commutator expectation is nonzero. 

However, the situation changes qualitatively for the state-dependent optimal relations derived in this paper.
For a non-faithful state, i.e., when $\lm=0$, our generalized relations~\eqref{main:UR} and~\eqref{main:UR2} reduce to the conventional Robertson and Schr\"odinger relations, respectively.
In what follows, we therefore restrict our attention to faithful states.
Moreover, since the maximally mixed state requires separate treatment, we further assume that $\rho$ is non-maximally mixed.
Let $P_{\min}$ and $P_{\max}$ denote the spectral projections of $\rho$ corresponding to $\lm$ and $\lM$, respectively.

\begin{proposition}
\label{prop:equality-gR}
Let $\rho$ be faithful and non-maximally mixed, and suppose that
\(
|\Exp{[A,B]}_\rho|>0
\).
Then equality holds in relation \eqref{main:UR} if and only if there exist $\alpha,\beta,t\in\mathbb R$, with $t\neq0$, and a nonzero operator $C$ satisfying
\begin{equation}\label{eq:C-extreme-support}
C=P_{\max}CP_{\min},
\end{equation}
such that
\begin{equation}\label{eq:equality-gR}
\begin{aligned}
A&=\alpha\I+C+C^\dagger,\\
B&=\beta\I+it(C-C^\dagger).
\end{aligned}
\end{equation}
\end{proposition}
The pair in Eq.~\eqref{eq:ex} is a simple example of the above equality construction.

{\it Proof.}
The sufficiency follows by direct substitution.
From Eq.~\eqref{eq:C-extreme-support}, we have
\begin{equation}
C^2=(C^\dagger)^2=0,
\quad
\rho CC^\dagger=\lM CC^\dagger,
\quad
\rho C^\dagger C=\lm C^\dagger C.
\end{equation}
Writing
$s:=\Tr(C^\dagger C)=\Tr(CC^\dagger)>0$
and using Eq.~\eqref{eq:equality-gR}, we obtain
\begin{align}
V_\rho(A)&=(\lM+\lm)s,\\
V_\rho(B)&=t^2(\lM+\lm)s,
\end{align}
and
\begin{equation}
\Exp{[A,B]}_\rho
=
-2it(\lM-\lm)s.
\end{equation}
These expressions saturate Eq.~\eqref{main:UR}.

For the necessity, we trace the equality conditions through the proof
of Theorem~\ref{thm1}.
Since $|\Exp{[A,B]}_\rho|>0$, both variances are nonzero.
Equality in Eq.~\eqref{eq:diagonal-reduction}, applied to the centered
observables $\hat A$ and $\hat B$, then requires their diagonal parts
to vanish:
\begin{equation}\label{eq:centered-diagonal-zero}
(\hat A)_d=(\hat B)_d=0.
\end{equation}
In what follows, let
$a_{ij}:=\langle i|\hat A|j\rangle$
and
$b_{ij}:=\langle i|\hat B|j\rangle$.

Next, equality must also hold when the spectral estimate
Eq.~\eqref{eq:wq3} is applied to the two factors in
Eq.~\eqref{eq:wq1}.
Since all terms in these factors are nonnegative, equality requires
Eq.~\eqref{eq:wq3} to be saturated for every pair $(i,j)$ for which
either $a_{ij}$ or $b_{ij}$ is nonzero.
For a faithful state, equality in Eq.~\eqref{eq:wq3} is equivalent to
\begin{equation}\label{eq:spectral-equality-condition}
\lambda_i\lM=\lambda_j\lm.
\end{equation}
Together with
$\lm\leq\lambda_i\leq\lambda_j\leq\lM$,
this is possible only when
\begin{equation}\label{eq:extremal-eigenvalue-pair}
\lambda_i=\lm,
\quad
\lambda_j=\lM.
\end{equation}
Hence, the only nonzero blocks of the centered observables are those
connecting the eigenspaces associated with $\lm$ and $\lM$.
Consequently,
\begin{align*}
\hat A
&=
P_{\max}\hat A P_{\min}
+
P_{\min}\hat A P_{\max},
\\
\hat B
&=
P_{\max}\hat B P_{\min}
+
P_{\min}\hat B P_{\max}.
\end{align*}
Since $\hat A$ and $\hat B$ are Hermitian, defining
\begin{equation}\label{eq:CD-extreme-blocks}
C:=P_{\max}\hat A P_{\min},
\quad
D:=P_{\max}\hat B P_{\min},
\end{equation}
gives
\begin{equation}
\hat A=C+C^\dagger,
\quad
\hat B=D+D^\dagger.
\end{equation}

Equality in the Cauchy--Schwarz inequality used in passing from
Eq.~\eqref{eq:wq4} to Eq.~\eqref{eq:wq1} can be characterized
explicitly.
Introducing
\begin{equation}
w_{ij}
:=
\frac{\lambda_j-\lambda_i}{\lambda_d-\lambda_1},
\end{equation}
there exists a single complex number $z\neq0$ such that
\begin{equation}
\sqrt{w_{ij}}\,\overline{b_{ij}}
=
z\sqrt{w_{ij}}\,\overline{a_{ij}}
\end{equation}
for every pair $i<j$.
By Eq.~\eqref{eq:extremal-eigenvalue-pair}, all nonzero matrix
elements occur for $\lambda_i=\lm$ and $\lambda_j=\lM$, for which
$w_{ij}=1$.
Therefore,
\begin{equation}
\overline{b_{ij}}
=
z\,\overline{a_{ij}}
\end{equation}
for all nonzero extremal matrix elements.
Using the definitions in Eq.~\eqref{eq:CD-extreme-blocks}, this is
precisely the operator relation
\begin{equation}\label{eq:D-zC}
D=zC.
\end{equation}

Finally, Eq.~\eqref{eq:wq4} was obtained from
Eq.~\eqref{eq:commutator-elements} by using
$|\Im S|\leq|S|$, where
\begin{equation}
S:=
\sum_{i<j}
\frac{\lambda_j-\lambda_i}{\lambda_d-\lambda_1}
a_{ij}\overline{b_{ij}}.
\end{equation}
Equality in Eq.~\eqref{eq:wq4} therefore requires $\Re S=0$.
Using Eqs.~\eqref{eq:extremal-eigenvalue-pair}
and~\eqref{eq:D-zC}, we obtain
$S=z\Tr(C^\dagger C)=zs$.
Since $s>0$, the condition $\Re S=0$ reduces to $\Re z=0$.
Hence $z=it$ for some $t\in\mathbb R\setminus\{0\}$.
Together with
$\hat A=C+C^\dagger$,
$\hat B=D+D^\dagger$,
and Eq.~\eqref{eq:D-zC}, this gives
\begin{equation}
\hat B
=
it(C-C^\dagger).
\end{equation}
Restoring the expectation-value parts of the centered observables
then yields Eq.~\eqref{eq:equality-gR}.
\qed

The covariance term enlarges this equality family in a precise manner for the generalized Schr\"odinger relation. 

\begin{proposition}
\label{prop:equality-gS}
Let $\rho$ be faithful and non-maximally mixed, and suppose that
$|\Exp{[A,B]}_\rho|>0$.
Then equality holds in relation~\eqref{main:UR2} if and only if there
exist $\alpha,\beta\in\mathbb R$, a nonzero operator
$C=P_{\max}CP_{\min}$, and a complex number $z$ with $\Im z\neq0$,
such that
\begin{equation}\label{eq:equality-gS}
\begin{aligned}
A&=\alpha\I+C+C^\dagger,\\
B&=\beta\I+zC+\overline{z}\,C^\dagger.
\end{aligned}
\end{equation}
\end{proposition}

{\it Proof.}
For the sufficiency, letting
$s:=\Tr(C^\dagger C)=\Tr(CC^\dagger)>0$
and using Eq.~\eqref{eq:equality-gS}, a computation analogous to that
in the proof of Proposition~\ref{prop:equality-gR} yields
\begin{align}
V_\rho(A)&=(\lM+\lm)s,\\
V_\rho(B)&=|z|^2(\lM+\lm)s,\\
{\rm Cov}_\rho(A,B)&=(\Re z)(\lM+\lm)s,
\end{align}
while
\begin{equation}
\Exp{[A,B]}_\rho
=
-2i(\Im z)(\lM-\lm)s.
\end{equation}
Substitution into Eq.~\eqref{main:UR2} shows that equality is attained.

For the necessity, assume equality in Eq.~\eqref{main:UR2}.
Since $|\Exp{[A,B]}_\rho|>0$, both variances are strictly positive.
Choose $u,v>0$ such that
\begin{equation}\label{eq:gS-balance}
u^2V_\rho(A)=v^2V_\rho(B),
\end{equation}
and, following the construction in Eq.~\eqref{PQ}, apply it to the
centered observables:
\begin{equation}\label{eq:PQ-centered}
P:=u\hat A-v\hat B,
\quad
Q:=u\hat A+v\hat B.
\end{equation}
Since $\hat A$ and $\hat B$ are centered, so are $P$ and $Q$.
Moreover,
\begin{equation}\label{eq:PQ-comm-equality}
[P,Q]=2uv[A,B],
\end{equation}
and hence
\begin{equation}
\Exp{[P,Q]}_\rho
=
2uv\,\Exp{[A,B]}_\rho\neq0.
\end{equation}

Using Eq.~\eqref{eq:gS-balance}, the same calculation as that leading
to Eq.~\eqref{URnormPQ2} gives
\begin{equation}\label{eq:PQ-product-equality}
V_\rho(P)V_\rho(Q)
=
4u^2v^2
\left[
V_\rho(A)V_\rho(B)
-
{\rm Cov}_\rho(A,B)^2
\right].
\end{equation}
On the other hand, Eq.~\eqref{eq:PQ-comm-equality} gives
\begin{equation}
\left|\Exp{[P,Q]}_\rho\right|^2
=
4u^2v^2
\left|\Exp{[A,B]}_\rho\right|^2.
\end{equation}
Therefore, equality in Eq.~\eqref{main:UR2} is equivalent, through
Eq.~\eqref{eq:PQ-product-equality}, to
\begin{equation}\label{eq:PQ-gR-equality}
V_\rho(P)V_\rho(Q)
=
\frac{(\lM+\lm)^2}{4(\lM-\lm)^2}
\left|\Exp{[P,Q]}_\rho\right|^2,
\end{equation}
namely, equality in the generalized Robertson relation~\eqref{main:UR} for the pair $P,Q$.

Since $P$ and $Q$ are centered and $|\Exp{[P,Q]}_\rho|>0$, Proposition~\ref{prop:equality-gR} implies that there exist a nonzero operator $E=P_{\max}EP_{\min}$ and a real number $t\neq0$ such that
\begin{equation}\label{eq:PQ-gR-form}
P=E+E^\dagger,
\quad
Q=it(E-E^\dagger).
\end{equation}
Solving Eq.~\eqref{eq:PQ-centered} for the centered observables gives
\begin{equation}\label{eq:AB-from-PQ}
\hat A=\frac{P+Q}{2u},
\quad
\hat B=\frac{Q-P}{2v}.
\end{equation}
Substitution of Eq.~\eqref{eq:PQ-gR-form} into
Eq.~\eqref{eq:AB-from-PQ} yields
\begin{align}
\hat A
&=
\frac{1+it}{2u}E
+
\frac{1-it}{2u}E^\dagger,
\label{eq:Ahat-from-E}
\\
\hat B
&=
\frac{-1+it}{2v}E
+
\frac{-1-it}{2v}E^\dagger.
\label{eq:Bhat-from-E}
\end{align}
Defining
\begin{equation}\label{eq:C-from-E}
C:=
\frac{1+it}{2u}E,
\end{equation}
we have $C=P_{\max}CP_{\min}$ and
\begin{equation}
\hat A=C+C^\dagger.
\end{equation}
Furthermore, Eq.~\eqref{eq:Bhat-from-E} can be written as
\begin{equation}
\hat B
=
zC+\overline{z}\,C^\dagger,
\end{equation}
where
\begin{equation}\label{eq:z-from-t}
z
=
\frac{u}{v}\,
\frac{-1+it}{1+it}.
\end{equation}
Since $u/v>0$ and $t\neq0$,
\begin{equation}
\Im z
=
\frac{u}{v}\,
\frac{2t}{1+t^2}
\neq0.
\end{equation}
Finally, restoring the expectation-value parts of the centered
observables yields Eq.~\eqref{eq:equality-gS}.
\qed

Propositions~\ref{prop:equality-gR} and~\ref{prop:equality-gS} reveal how the covariance term changes the equality structure of the optimized bounds.
Writing $z=x+iy$ in Eq.~\eqref{eq:equality-gS}, one has
\begin{equation}
\hat B
=
x(C+C^\dagger)
+
iy(C-C^\dagger).
\end{equation}
The real part $x$ contributes to the symmetric covariance, while the imaginary part $y$ contributes to the commutator expectation.
Saturation of the generalized Robertson relation is recovered as the special case $x=0$, whereas the covariance term in the generalized Schr\"odinger relation allows the full nonreal family $z\in\mathbb C\setminus\mathbb R$.

This comparison provides a physical interpretation of the improvement that remains within the preparation-uncertainty framework.
For a faithful mixed state, neither the conventional Robertson relation nor the conventional Schr\"odinger relation can be tight for a genuinely noncommuting pair with nonzero commutator expectation.
By contrast, the state-dependent optimal coefficient restores genuine noncommuting saturation.
The observables that realize it are not arbitrary: their centered parts couple only the eigenspaces associated with the smallest and largest eigenvalues of $\rho$.
Thus, the optimized relations identify precisely the observable configurations for which the lower bound becomes an exact expression of the uncertainty trade-off in the prescribed mixed state.

For qubits this structure has a particularly simple geometric form. 
In what follows, we denote vectors in $\mathbb{R}^3$ by
${\bm a}$, ${\bm b}$, and ${\bm n}$, and write the vector of Pauli matrices as
${\bm \sigma}=(\sigma_x,\sigma_y,\sigma_z)$.
\begin{corollary}
\label{cor:equality-qubit}
Let $\rho$ be a faithful, non-maximally mixed qubit state, i.e.,
\begin{equation}\label{eq:BlochStates}
\rho
=
\frac{1}{2}\bigl(\I+r\,\bm n\cdot\bm\sigma\bigr),
\quad
0<r<1,
\quad
|\bm n|=1,
\end{equation}
and consider spin observables in the directions $\bm a$ and $\bm b$,
with $|\bm a|=|\bm b|=1$:
\begin{equation}\label{spin}
A=\bm a\cdot\bm\sigma,
\quad
B=\bm b\cdot\bm\sigma.
\end{equation}

For the conventional uncertainty relations, Robertson's relation
\eqref{eq:Rob} is never saturated, whereas Schr\"odinger's relation
\eqref{eq:SUR} is saturated if and only if
\begin{equation}
\bm b=\pm\bm a.
\end{equation}

Under the additional condition
$|\Exp{[A,B]}_\rho|>0$,
the generalized Schr\"odinger relation~\eqref{main:UR2} is saturated
if and only if
\begin{equation}
\bm a\perp\bm n,
\quad
\bm b\perp\bm n.
\end{equation}
The generalized Robertson relation~\eqref{main:UR} is saturated if
and only if, in addition,
\begin{equation}
\bm a\perp\bm b.
\end{equation}
\end{corollary}

{\it Proof.}
We first consider the conventional relations.
For a normalized spin observable $A=\bm a\cdot\bm\sigma$,
\begin{equation}
V_\rho(A)
=
1-r^2(\bm a\cdot\bm n)^2
>0,
\end{equation}
since $0<r<1$, and similarly $V_\rho(B)>0$.
As shown above, saturation of the conventional Robertson relation for
a faithful, non-maximally mixed state necessarily implies
$\Exp{[A,B]}_\rho=0$.
Its right-hand side then vanishes, whereas
$V_\rho(A)V_\rho(B)>0$.
Hence the conventional Robertson relation cannot be saturated.

For the conventional Schr\"odinger relation, the equality condition
\eqref{eq:standard-SR-equality} gives
\begin{equation}
\hat B=t\hat A
\end{equation}
for some $t\in\mathbb R\setminus\{0\}$.
Comparing the Pauli-vector parts yields
$\bm b=t\bm a$.
Since $|\bm a|=|\bm b|=1$, this is equivalent to
\begin{equation}
\bm b=\pm\bm a.
\end{equation}
Conversely, either of these two choices makes the centered observables
real-linearly dependent and therefore saturates the conventional
Schr\"odinger relation.

We next consider the generalized relations.
For a faithful, non-maximally mixed qubit state, both extremal spectral projections are rank one, and hence $P_{\max}{\cal B}(\mathcal H)P_{\min}$ is one-dimensional over $\mathbb C$.
A centered observable has nonzero blocks only between these two eigenspaces precisely when its Bloch direction is perpendicular to $\bm n$.
Therefore, Proposition~\ref{prop:equality-gS} implies that the generalized Schr\"odinger relation is saturated precisely when
\begin{equation}
\bm a\perp\bm n,
\quad
\bm b\perp\bm n,
\end{equation}
where the assumption $|\Exp{[A,B]}_\rho|>0$ excludes parallel directions.

Under these conditions, the extremal blocks associated with $\bm a$ and $\bm b$ differ by a complex proportionality constant.
By Proposition~\ref{prop:equality-gR}, saturation of the generalized Robertson relation further requires this proportionality constant to be purely imaginary, which for qubit Bloch vectors is equivalent to
\begin{equation}
\bm a\perp\bm b.
\end{equation}
This proves the stated conditions.
\qed

\begin{remark}
For qubit spin observables, the condition
$\bm a\perp\bm b$ means that the eigenbases of $A$ and $B$
are mutually unbiased.
Hence, the equality condition for the generalized Robertson relation
shows that mutually unbiased observables can saturate the bound,
provided that both Bloch directions are perpendicular to the Bloch
vector $\bm n$ of the state.
For example, when $\bm n$ is along the $z$ direction,
the pair $A=\sigma_x$ and $B=\sigma_y$ saturates the generalized
Robertson relation.
This is in sharp contrast with the conventional Robertson relation,
which cannot be saturated for a faithful mixed qubit state.
\end{remark}
\end{gkblock}

\section{Comparison with Robertson and Schr\"odinger bounds}\label{Sec:4} 
\begin{figure*}[th]
    \centering
\includegraphics[width=0.8\textwidth]{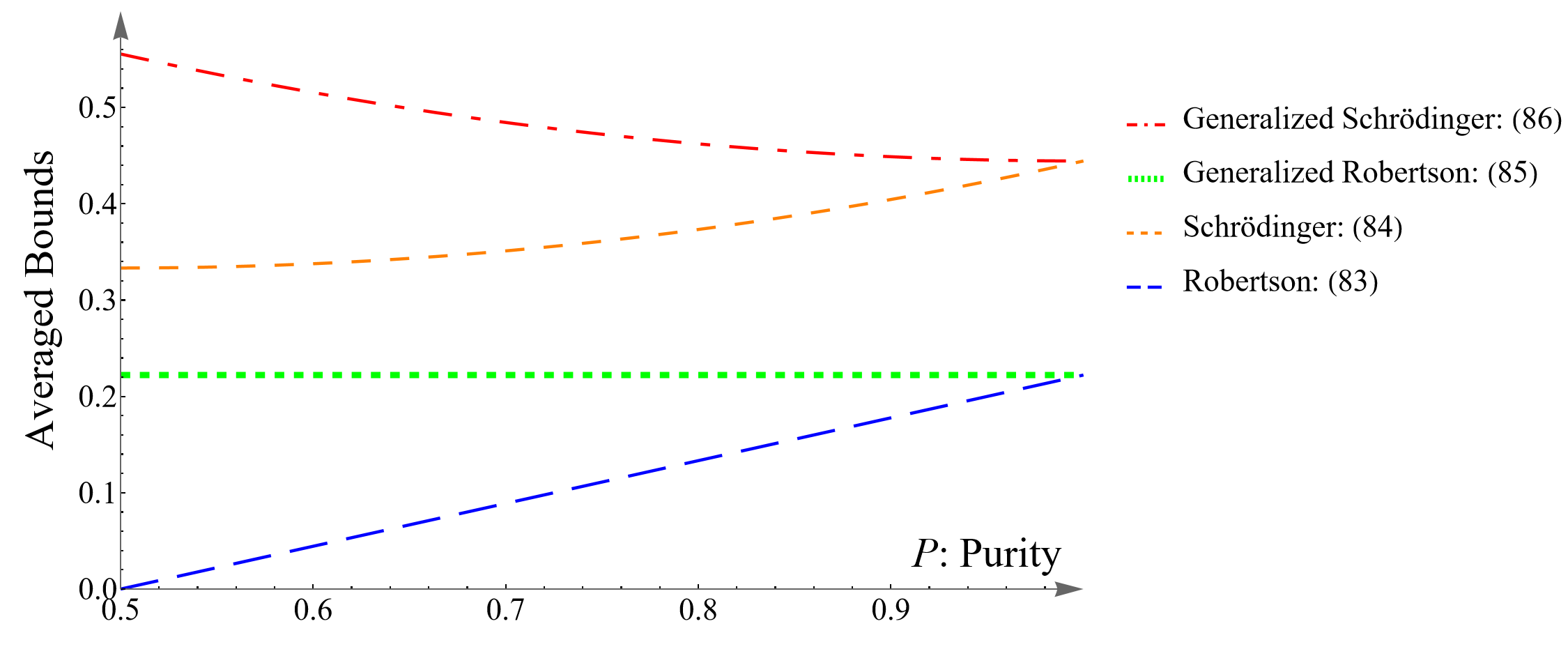} 
    \caption{Comparison of the averaged bounds of Robertson's relation~\eqref{rob2} (blue, long-dashed), Schr\"odinger's relation~\eqref{sch2} (orange, short-dashed), the generalized Robertson's relation~\eqref{grob2} (green, dotted), and the generalized Schr\"odinger's relation~\eqref{gsch2} (red, dot-dashed).  
    }
    \label{fig:comp}
\end{figure*}

In this section, we compare our uncertainty relations~\eqref{main:UR} and~\eqref{main:UR2} with the Robertson and Schr\"odinger relations in the qubit system.
Throughout this section, we restrict our attention to non-maximally mixed
states, for which the purity satisfies $1/2<P\leq1$.
To ensure an unbiased and consistent comparison, we average the respective bounds over all spin observables $A$ and $B$, thereby eliminating any observable-dependent bias.
Specifically, letting $A,B$ be (normalized) spin observables \eqref{spin}, we compute the averaged bounds over all unit vectors $ \bm{a}$, $\bm{b}$ uniformly distributed on the sphere (See \cite{MKHC2} for the details). 
The averaged bounds (i.e., the right-hand sides) of Robertson's relation~\eqref{eq:Rob} and Schr\"odinger's relation~\eqref{eq:SUR} are given by 	\begin{align}
		&\langle B_{\rm R}(A,B,\rho)\rangle_{\rm av} := \Bigl\langle \frac{1}{4} |\langle [A,B]\rangle_\rho|^2\Bigr\rangle_{\rm av} = \frac{2}{9}(2P-1) \label{rob2}\\
		&\langle B_{\rm S}(A,B,\rho)\rangle_{\rm av}:= \Bigl\langle \frac{1}{4} |\langle [A,B]\rangle_\rho|^2\Bigr\rangle_{\rm av} + \Bigl\langle {\rm Cov}_\rho(A,B)^2 \Bigr\rangle_{\rm av} \nonumber \\ 
        &= \frac{2}{9}(2P-1) + \frac{2}{9}(2P^2-4P+3) = \frac{4}{9}(P^2-P+1),\label{sch2}
	\end{align}
where \( P = \Tr\rho^2 \) denotes the purity of the quantum state \( \rho \). 
Since the eigenvalues of a qubit state $\rho$ are given by \(\lambda_{\max} = (1 + \sqrt{2P-1})/2, \  \lambda_{\min} = (1 - \sqrt{2P-1})/2 \), one has \(\frac{(\lM + \lm)^2}{(\lM - \lm)^2} = \frac{1}{2P-1}\). With this modification, the averaged bounds of the generalized Robertson's relation~\eqref{main:UR} and the generalized Schr\"odinger's relation~\eqref{main:UR2} become
\begin{align}
&\langle B_{\rm gR}(A,B,\rho)\rangle_{\rm av} =\frac{2}{9},\label{grob2} \\
&\langle B_{\rm gS}(A,B,\rho)\rangle_{\rm av} =\frac{4}{9}(P^2 - 2P + 2). \label{gsch2} 
\end{align}
Fig.~\ref{fig:comp} shows the averaged bounds~\eqref{rob2}, \eqref{sch2}, \eqref{grob2}, and~\eqref{gsch2} plotted as functions of the purity \( P \) in the range \(1/2<P\le 1\). 
It is evident that both the generalized Robertson and Schr\"odinger relations provide tighter bounds as the quantum state becomes more mixed.
It is particularly interesting to note that, whereas the original Robertson bound fails to capture the trade-off in the highly mixed regime, our bound~\eqref{main:UR} consistently reflects a nontrivial trade-off regardless of the state's purity.

\section{Applications to measurement type uncertainty relations}\label{Sec:5} 

In this section, we show that measurement-type uncertainty relations whose derivations rely on Robertson's inequality can likewise be strengthened by using our improved relation in its place.
Among the best-known formulations of measurement-type uncertainty relations that treat measurement error and disturbance in a quantitative manner via root-mean-square deviations are the Arthurs--Goodman inequality~\cite{ArthursGoodman1988} and Ozawa’s inequalities~\cite{Ozawa2003,Ozawa2004}. 

Let \( \rho \) and \( \rho_{\mathrm{app}} \) be the initial states of the system of interest and the apparatus, respectively.  
Let \( A_{\mathrm{app}} \) and \( B_{\mathrm{app}} \) be the corresponding commuting meter observables.  
The errors for \( X = A, B \) are defined as  
\(
\varepsilon(X) := \sqrt{\Tr\left[(\rho \otimes \rho_{\mathrm{app}})\, N_X^2 \right]},
\)  
where  
\(
N_X := U^{\dagger}(\mathbb{I} \otimes X_{\mathrm{app}})U - X \otimes \mathbb{I}.
\)  
\gk{Under the unbiasedness assumption, the Arthurs--Goodman inequality reads}
\begin{equation}\label{AGinew}
\varepsilon(A)\varepsilon(B)
\ge \frac{1}{2} \left| \langle [A, B] \rangle_{\rho} \right|.
\end{equation}
\gk{Without the unbiasedness assumption, Ozawa established the more general relation:}
\begin{equation}\label{Ozawa1}
\varepsilon(A)\varepsilon(B)
+
\varepsilon(A)\sigma(B)
+
\sigma(A)\varepsilon(B)
\ge
\frac{1}{2} \left| \langle [A, B] \rangle_{\rho} \right|,
\end{equation}
where $\sigma(X):= \sqrt{V_\rho(X)}$ denotes the standard deviation of $X$. 
\gk{Defining the root-mean-square disturbance by $\eta(B):=\sqrt{\Tr\bigl[(\rho\otimes\rho_{\mathrm{app}})\bigl(U^{\dagger}(B\otimes\mathbb I)U-B\otimes\mathbb I \bigr)^{2}\bigr]}$, Ozawa's error--disturbance relation reads}
\begin{equation}\label{Ozawa2}
\varepsilon(A)\eta(B)
+
\varepsilon(A)\sigma(B)
+
\sigma(A)\eta(B)
\ge
\frac{1}{2} \left| \langle [A, B] \rangle_{\rho} \right|. 
\end{equation}
\gk{
The proofs of these measurement-type inequalities rely directly on applications of the Robertson-type uncertainty relation \cite{ArthursGoodman1988,Ozawa2003,Ozawa2004}. 
To make the spectral improvement transparent, let us introduce the joint initial state
\(\Omega=\rho\otimes\rho_{\mathrm{app}}\).
Let $\lambda_{\max}$ and $\lambda_{\min}$ denote the largest and
smallest eigenvalues of $\rho$, respectively, and let
$\mu_{\max}$ and $\mu_{\min}$ denote those of
$\rho_{\mathrm{app}}$. 
The largest and smallest eigenvalues of $\Omega$ are then given by
\(\Lambda_{\max}=\lambda_{\max}\mu_{\max}, \Lambda_{\min}=\lambda_{\min}\mu_{\min}.\)
Define the prefactor associated with the generalized Robertson relation~\eqref{main:UR} on the joint system by
\begin{equation}
\kappa_\Omega:=\frac{\Lambda_{\max}+\Lambda_{\min}}{(\Lambda_{\max}-\Lambda_{\min})}=\frac{\lambda_{\max}\mu_{\max}+\lambda_{\min}\mu_{\min}}{(\lambda_{\max}\mu_{\max}-\lambda_{\min}\mu_{\min})}.
\end{equation}
Replacing the Robertson-type estimate used in the derivations of the Arthurs--Goodman and Ozawa inequalities with the strengthened estimate provided by Theorem~1 yields the following generalized Arthurs--Goodman inequality:
\begin{equation}
\varepsilon(A)\varepsilon(B)
\ge \frac{\kappa_\Omega}{2}\left|\Exp{[A,B]}_\rho\right|.
\end{equation}
Similarly, the corresponding Ozawa-type inequalities become
\begin{align}
    \varepsilon(A)\varepsilon(B)
    &{}+\varepsilon(A)\sigma(B)+\sigma(A)\varepsilon(B)\ge \frac{\kappa_\Omega}{2}\left|\Exp{[A,B]}_\rho\right|,
\end{align}
\begin{align}
    \varepsilon(A)\eta(B)
    &{}+\varepsilon(A)\sigma(B)+\sigma(A)\eta(B) \ge \frac{\kappa_\Omega}{2}\left|\Exp{[A,B]}_\rho\right|.\label{eq:genetralized-ozawa-error-disturbance}    
\end{align}
These generalized inequalities replace the conventional constant \(1/2\) with a coefficient determined by the eigenvalues of both the system and apparatus states, thereby yielding tighter lower bounds.  
In the special case where the apparatus state \(\rho_{\mathrm{app}}\) is non-faithful, particularly in the pure-state case, these relations reduce to the standard Arthurs--Goodman and Ozawa inequalities.}

\gk{It is instructive to compare these generalized inequalities with Branciard's error--trade-off relation~\cite{Branciard2013}. 
The generalized inequalities in this section are obtained by applying Theorem~1, rather than the standard Robertson relation, to the relevant operators on the joint initial state \(\Omega=\rho\otimes\rho_{\mathrm{app}}\): the noise operators in the Arthurs--Goodman derivation, and the noise and disturbance operators in the Ozawa derivations.   
By contrast, Branciard's relation is based on a different geometric argument.
Therefore, our result should not be read as a direct improvement of Branciard's bound.   
A normalized qubit comparison with Branciard's relation, including the definitions of the normalized errors, is given in Appendix~\ref{app:branciard}.
}

\section{CONCLUSIONS AND DISCUSSION}\label{Sec:6} 
In this paper, we have investigated a generalization of Robertson's uncertainty relation of the form given in Eq.~\eqref{II} and established the optimal bound in Eq.~\eqref{main:UR}.
For finite-dimensional systems and faithful states, this relation provides a strict improvement over the original Robertson bound and cannot be further improved. In this sense, it is noteworthy that the optimal bound depends solely on the smallest and largest eigenvalues of the quantum state \( \rho \), whereas in the case of~\eqref{I}, the optimal bound is determined by the smallest and second-smallest eigenvalues of \( \rho \). 
\gk{By incorporating the covariance term, we have further established relation~\eqref{main:UR2}, which provides a tight generalization of Schr\"odinger's uncertainty relation. Moreover, applying the same spectral improvement to the Robertson-type estimates used in the derivations of the Arthurs--Goodman and Ozawa inequalities yields corresponding generalizations of these measurement-type uncertainty relations.}

\gk{
We have also characterized the equality conditions of the optimized Robertson and Schr\"odinger relations.  For faithful mixed states, the conventional Robertson and Schr\"odinger bounds cannot be saturated by genuinely noncommuting observables with a nonzero commutator expectation.  In contrast, the optimized state-dependent relations admit noncommuting saturating pairs whose centered parts couple the eigenspaces associated with the smallest and largest eigenvalues of the state.  Thus the spectral refinement not only strengthens the numerical lower bound, but also restores genuine tightness of the uncertainty trade-off for mixed states.
}

\gk{Finally, and importantly, our relations are not merely mathematical
refinements of the conventional uncertainty bounds.
Their lower bounds retain the operationally transparent form of the
Robertson and Schr\"odinger relations, being expressed in terms of
experimentally accessible expectation values of observables.
Thus, the state-dependent improvement obtained here can, in principle,
be directly tested experimentally rather than serving only as an
abstract mathematical bound.}

\gk{More than a century after the advent of quantum mechanics, it is remarkable that a relation as fundamental as Robertson’s uncertainty relation still admits a tight state-dependent refinement.}

\section*{Acknowledgments}

We thank Kenta Koshihara, Kenjiro Yanagi, and Jaeha Lee for their valuable comments and discussions. G.K. was supported by JSPS KAKENHI Grant No.~24K06873.

\appendix

\section{\MakeUppercase{Comparison with Branciard's error--tradeoff relation}}
\begin{gkblock}
\label{app:branciard}

In this appendix, we compare Ozawa's inequality~\eqref{Ozawa2} and its
generalized form~\eqref{eq:genetralized-ozawa-error-disturbance} with
Branciard's error--tradeoff relation~\cite{Branciard2013}.
After presenting the comparison in normalized variables, we illustrate
it explicitly for a qubit system measured with a qubit apparatus.
The purpose is not to claim a general improvement over Branciard's
relation, but rather to clarify how our result strengthens Ozawa's
inequality and how this strengthened bound compares with Branciard's
geometric relation.

Using the normalized error and disturbance variables
\begin{equation}
x:=\frac{\epsilon(A)}{\sigma(A)},
\quad
y:=\frac{\eta(B)}{\sigma(B)},
\end{equation}
Branciard's error--disturbance relation takes the form
\begin{equation}
x^2+y^2+2\sqrt{1-C^2}\,xy\geq C^2,
\label{eq:branciard-normalized}
\end{equation}
where $C$ denotes the normalized commutator term,
\begin{equation}
C:=
\frac{\left|\Exp{[A,B]}_{\rho}\right|}
{2\sigma(A)\sigma(B)}.
\label{eq:C-equals-r}
\end{equation}
In terms of the same normalized variables, the standard Ozawa
inequality~\eqref{Ozawa2} becomes
\begin{equation}
xy+x+y\geq C,
\label{eq:ozawa-normalized}
\end{equation}
whereas the generalized Ozawa-type
inequality~\eqref{eq:genetralized-ozawa-error-disturbance} becomes
\begin{equation}
xy+x+y\geq \kappa_{\Omega} C.
\label{eq:gozawa-normalized}
\end{equation}

We now specialize to the case where both the system and the measuring
apparatus are qubits.
Consider the mutually unbiased spin observables
\begin{equation}\label{eq:pauli-example}
A=\sigma_x,
\quad
B=\sigma_y.
\end{equation}
Let the system state be
\begin{equation}\label{eq:branciard-system-state}
\rho
=
\frac{1}{2}\left(\I+r\sigma_z\right),
\quad
0<r\leq1,
\end{equation}
and denote its purity by
\begin{equation}
P:=\Tr\rho^2
=
\frac{1+r^2}{2}.
\end{equation}
Thus,
\begin{equation}
r=\sqrt{2P-1}.
\end{equation}
For the observables~\eqref{eq:pauli-example}, we have
$\sigma(A)=\sigma(B)=1$ and
\begin{equation}
C
=
\sqrt{2P-1}.
\label{eq:C-purity}
\end{equation}

Similarly, let the apparatus state be
\begin{equation}\label{eq:branciard-apparatus-state}
\rho_{\rm app}
=
\frac{1}{2}\left(\I+s\sigma_z\right),
\quad
0<s\leq1,
\end{equation}
and denote its purity by
\begin{equation}
P_{\rm app}
:=
\Tr\rho_{\rm app}^2
=
\frac{1+s^2}{2}.
\end{equation}
Hence,
\begin{equation}
s=\sqrt{2P_{\rm app}-1}.
\end{equation}

For the joint initial state
$\Omega=\rho\otimes\rho_{\rm app}$,
the spectral factor appearing in
Eq.~\eqref{eq:gozawa-normalized} is
\begin{equation}
\kappa_{\Omega}
=
\frac{1+rs}{r+s}.
\end{equation}
In terms of the purities, this becomes
\begin{equation}\label{eq:kappa-purity}
\kappa_{\Omega}
=
\frac{
1+\sqrt{(2P-1)(2P_{\rm app}-1)}
}{
\sqrt{2P-1}
+
\sqrt{2P_{\rm app}-1}
}.
\end{equation}
Equivalently,
\begin{equation}
\kappa_{\Omega}-1
=
\frac{
\left(1-\sqrt{2P-1}\right)
\left(1-\sqrt{2P_{\rm app}-1}\right)
}{
\sqrt{2P-1}
+
\sqrt{2P_{\rm app}-1}
}
\geq0.
\end{equation}
Therefore,
\begin{equation}
\kappa_{\Omega}\geq1,
\end{equation}
with equality if and only if either
$P=1$ or $P_{\rm app}=1$, namely, if either the system or the
apparatus state is pure.
Thus, the generalized relation~\eqref{eq:gozawa-normalized} is at
least as strong as the standard Ozawa relation and provides a strict
improvement whenever both the system and the apparatus states are
mixed.

For the present choice of observables, the three relations can be
written entirely in terms of the system purity.
Using Eq.~\eqref{eq:C-purity}, Branciard's relation becomes
\begin{equation}\label{eq:branciard-purity}
x^2+y^2
+
2\sqrt{2(1-P)}\,xy
\geq
2P-1,
\end{equation}
while the standard Ozawa relation becomes
\begin{equation}\label{eq:ozawa-purity}
xy+x+y
\geq
\sqrt{2P-1}.
\end{equation}
Our generalized Ozawa-type relation reads
\begin{equation}\label{eq:gozawa-purity}
xy+x+y
\geq
\kappa_{\Omega}\sqrt{2P-1},
\end{equation}
where $\kappa_{\Omega}$ is given by
Eq.~\eqref{eq:kappa-purity}.

Figure~\ref{fig:branciard-comparison} illustrates the comparison among
Branciard's relation~\eqref{eq:branciard-purity}, Ozawa's
relation~\eqref{eq:ozawa-purity}, and its generalized form
\eqref{eq:gozawa-purity}.
The system purity is fixed at
\begin{equation}
P=0.625,
\end{equation}
corresponding to $r=0.50$, while the apparatus purities are chosen as
\begin{equation}
P_{\rm app}
=
0.78125,
\quad
0.625,
\quad
0.53125,
\end{equation}
corresponding respectively to
$s=0.75,0.50,0.25$.

\begin{figure*}[t]
    \centering
    \begin{tabular}{ccc}
        \includegraphics[width=0.31\textwidth]{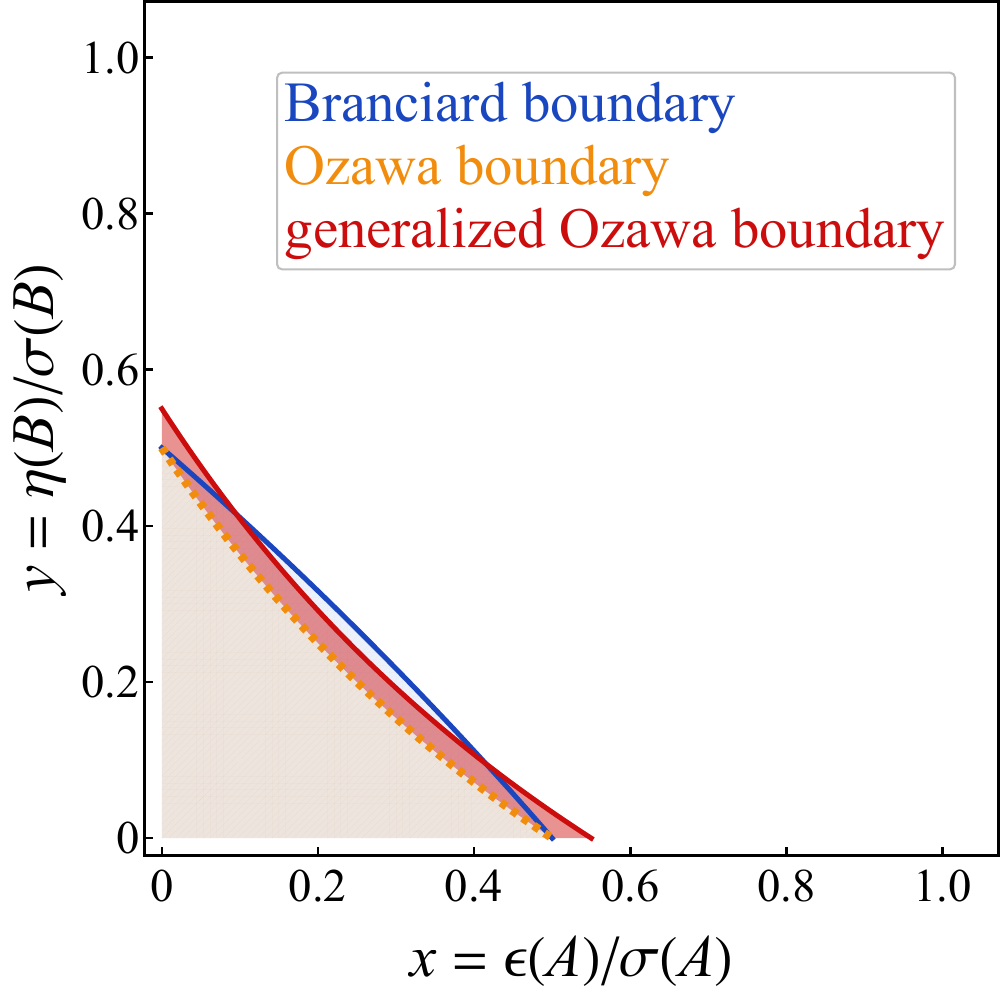}
        &
        \includegraphics[width=0.31\textwidth]{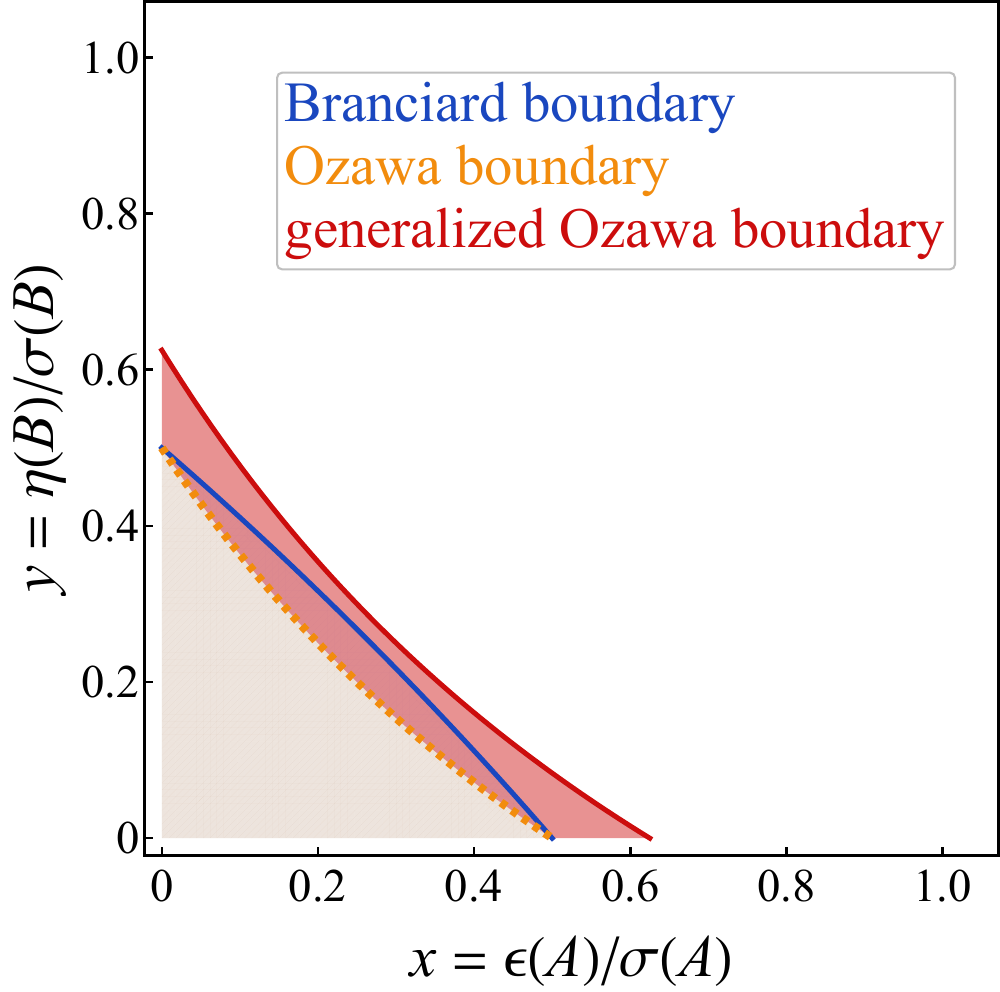}
        &
        \includegraphics[width=0.31\textwidth]{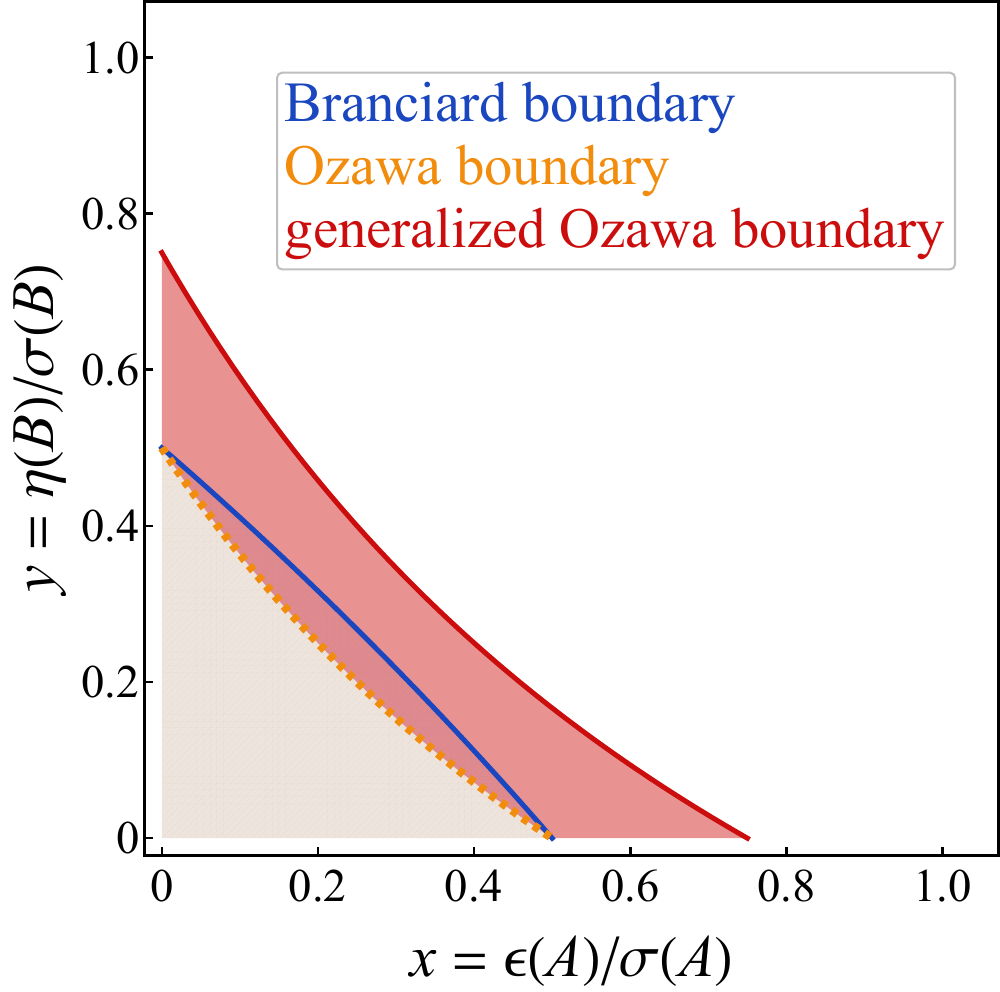}
        \\
        (a) $P=0.625$, $P_{\rm app}\simeq0.781$, $\kappa_{\Omega}=1.100$
        &
        (b) $P=0.625$, $P_{\rm app}=0.625$, $\kappa_{\Omega}=1.250$
        &
        (c) $P=0.625$, $P_{\rm app}\simeq0.531$, $\kappa_{\Omega}=1.500$
    \end{tabular}
    \caption{\begin{gkblock}
    Comparison of normalized error--disturbance regions for
    $A=\sigma_x$ and $B=\sigma_y$, with the system purity fixed at
    $P=0.625$.
    The solid blue curve represents Branciard's relation
    \eqref{eq:branciard-purity}, the dashed orange curve represents
    Ozawa's relation~\eqref{eq:ozawa-purity}, and the solid red curve
    represents the generalized Ozawa-type relation
    \eqref{eq:gozawa-purity}.
    The apparatus purities are
    $P_{\rm app}\simeq0.781$, $0.625$, and $0.531$ in panels
    (a), (b), and (c), respectively.
    The shaded regions are forbidden by the corresponding inequalities.
    In particular, the red shaded region is allowed by the standard
    Ozawa inequality but excluded by the generalized Ozawa-type
    inequality.
    As the apparatus purity decreases, the apparatus state becomes
    more mixed and $\kappa_{\Omega}$ increases, thereby enlarging the
    additional forbidden region.\end{gkblock}}
    \label{fig:branciard-comparison}
\end{figure*}

The corresponding improvement factors are $\kappa_{\Omega}=1.100,1.250,1.500$, respectively.
The shaded regions are forbidden regions, namely, regions that violate the corresponding inequalities, so that a larger forbidden region represents a stronger constraint.
In each panel, the additional region allowed by the standard Ozawa inequality but forbidden by the generalized Ozawa-type inequality is characterized by
\begin{equation}
\sqrt{2P-1}
\leq
xy+x+y
<
\kappa_{\Omega}\sqrt{2P-1}.
\end{equation}
The three panels show that, for a fixed system state, the spectral
refinement systematically strengthens the Robertson-based Ozawa bound as the apparatus state becomes more mixed.

Finally, we emphasize that the comparison with Branciard's relation should be interpreted with care.
The generalized Ozawa-type inequality is obtained by replacing the Robertson-type step in the derivation of Ozawa's inequality with
Theorem~\ref{thm1}.
Hence, it is a spectral refinement of a Robertson-based error--disturbance relation.
By contrast, Branciard's relation is derived from a different geometric argument and imposes a constraint of a qualitatively different form.
In particular, Branciard's relation is tight for pure states in the corresponding optimization setting, whereas the spectral improvement developed here becomes strict only when both the system and the apparatus states are mixed.
Therefore, there is no simple general ordering between the spectral refinement in Eq.~\eqref{eq:gozawa-purity} and Branciard's geometric bound in Eq.~\eqref{eq:branciard-purity}.
Combining the state-spectral improvement developed in this work with Branciard's geometric framework remains an interesting open problem.
\end{gkblock}

\bibliographystyle{apsrev4-2}
\bibliography{ref_UR}

\end{document}